\documentclass{article}%
\usepackage{amsfonts}
\usepackage{amsmath}
\usepackage{amssymb}
\usepackage{graphicx}%
\providecommand{\U}[1]{\protect\rule{.1in}{.1in}}

\begin{document}

\title{Self-Diffusion Coefficients of Lennard-Jones Liquids and Gases for Various
Models in the Modified Free Volume Theory:\ Tables}
\author{Jagtar Singh Hunjan and Byung Chan Eu\\Department of Chemistry, McGill University,\\810 Sherbrooke St. West, Montreal, QC H3A 2K6 Canada}
\maketitle

\begin{abstract}
We present tables for self-diffusion coefficients of the Lennard-Jones liquids
and gases for various model formulas in the modified free volume theory of
diffusion in the case of reduced temperatures $T^{\ast}=6.0$, $4.0$, $1.3$,
$1.0$, $0.8$, $0.7$ with reduced density ranging from $\rho^{\ast}=0.025$ to
$1$. Their accuracies are compared with the molecular dynamic simulation
results. In the gas and liquid density regimes, the formula $D_{CERC}$ for the
self-diffusion coefficient is accurate within maximum 4\% of the molecular
dynamics simulation values reported in the literature. It is the most reliable
among the formulas presented.

\end{abstract}

\section{Self-Diffusion Coefficients of the Lennard-Jones Fluids}

In this note, the self-diffusion coefficients are tabulated for Lennard-Jones
(LJ) fluids, which are calculated with various model formulas in the modified
free volume (MFV) theory\cite{mfv}. The reduced temperatures studied are
$T^{\ast}=6.0$, $4.0$, $1.3$, $1.0$, $0.8$, $0.7$ and the reduced density
ranges from $\rho^{\ast}=0.025\ $to $\rho^{\ast}=1$. \ Here $T^{\ast}%
=k_{B}T/\epsilon$ and $\rho^{\ast}=\rho\sigma^{3}$, where $\epsilon$ is the
well depth and $\sigma$ is the contact diameter of the LJ fluid, $k_{B}$, $T$,
and $\rho$ denoting, respectively, the Boltzmann constant, temperature, and
number density. The various formulas in the MFV theory of diffusion computed
in this article are modifications of the original form for the self-diffusion
coefficient of simple fluids in the MFV theory\cite{mfv}:%
\begin{equation}
D=\frac{3}{8\rho\sigma^{2}}\sqrt{\frac{k_{B}T}{{\pi}m}}\exp(-\alpha
v_{0}/{v_{f}}), \label{1}%
\end{equation}
where ${v_{f}}$ is the mean free volume, $v_{0}$ is the critical volume, and
$\alpha$ is the free volume overlap parameter. In the MFV\ theory, the mean
free volume $v_{f}$ is given by the generic van der Waals (GvdW) equation of
state\cite{GvdW}, which was originally identified with the formula%
\begin{equation}
v_{f}=v\left[  1-\frac{2\pi}{3}\beta\rho{\int_{0}^{\sigma}}dr{r^{3}}\frac
{du}{dr}g(r,\rho,T)\right]  ^{-1} \label{2}%
\end{equation}
with $u$ denoting the intermolecular interaction potential, e.g., the LJ
potential; $g(r,\rho,T)$ the pair correlation function (radial distribution
function); and $\beta=1/k_{B}T$. The critical volume was originally taken
$v_{0}=\pi\sigma^{3}/6$, the covolume of hard spheres of diameter $\sigma$.
The overlap parameter was chosen suitably, usually slightly less than unity,
as an adjustable parameter. In the following, this basic formula is suitably
modified such that the self-diffusion coefficient is expressible in terms of
molecular parameters only, yet in sufficiently good accuracy for the
self-diffusion coefficient.

\subsection{Model $D_{HS}$}

With the definition of $r^{\ddagger}$ as the maximum position of the function%
\begin{equation}
I\left(  r\right)  =-r^{3}\frac{du}{dr}g(r,\rho,T), \label{3}%
\end{equation}
that is,%
\begin{equation}
\left(  {\frac{dI}{dr}}\right)  _{r=r^{\ddagger}}=-\left[  \frac{d}{dr}%
r^{3}\frac{du}{dr}g(r,\rho,T)\right]  _{r=r^{\ddag}}=0, \label{4}%
\end{equation}
the mean free volume $v_{f}$ is modified to the formula%
\begin{equation}
v_{f}=v\left[  1-\frac{2\pi}{3}\beta\rho{\int_{0}^{r^{\ddagger}}}dr{r^{3}%
}\frac{du}{dr}g(r,\rho,T)\right]  ^{-1} \label{5}%
\end{equation}
and at the same time $\alpha v_{0}$ in Eq. (\ref{1}) is set%
\begin{equation}
\alpha v_{0}\Rightarrow\frac{\pi}{6}r^{\ddagger3}\equiv v^{c}. \label{6}%
\end{equation}
Thus the self-diffusion coefficient takes the form\cite{laghaei}%
\begin{equation}
D_{HS}=\frac{3}{8\rho\sigma^{2}}\sqrt{\frac{k_{B}T}{{\pi}m}}\exp(-{v^{c}%
}/{v_{f}}), \label{7}%
\end{equation}
which is now free from adjustable parameters. It improves the accuracy of the
self-diffusion coefficient considerably.

\subsection{Model $D_{HSRC}$}

The diameter $r^{\ddagger}$ a little exaggerates the excluded volume for the
LJ fluid. To rectify this feature the diameter of the excluded volume is
modified by the following formula\cite{laghaei}%
\begin{equation}
r_{c}=r^{\ddag}-\left\vert r_{fm}-r_{pm}\right\vert \simeq r^{\ddag
}-0.1222\sigma. \label{8}%
\end{equation}
Here $r_{fm}$ is the minimum position of $I\left(  r\right)  $ and $r_{pm}$ is
the minimum of the potential energy or the zero of the intermolecular force.
If the free volume overlap parameter $\alpha$ is taken as $\alpha=\left(
r_{c}/r^{\ddag}\right)  ^{3}$, which may be interpreted as the ratio of the
volume of a sphere of diameter $r_{c}$ to that of a sphere of diameter
$r^{\ddag}$, then the critical volume turns out to be simply equal to%
\begin{equation}
\alpha v_{0}\Rightarrow\frac{\pi}{6}r_{c}^{3}\equiv v^{0}. \label{9}%
\end{equation}
We keep the mean free volume $v_{f}$ as in Eq. (\ref{5}). With Eq. (\ref{9})
for the critical volume facilitating diffusion the self-diffusion formula now
takes the form\cite{excluded}%
\begin{equation}
D_{HSRC}=\frac{3}{8\rho\sigma^{2}}\sqrt{\frac{k_{B}T}{{\pi}m}}\exp(-{v^{0}%
}/{v_{f}}), \label{10}%
\end{equation}
which is free from an adjustable parameter.

\subsection{Model $D_{CE}$}

The pre-exponential factor in the self-diffusion coefficient formulas given
earlier represents the self-diffusion coefficient of a hard sphere fluid.
However, it appears to be a poor approximation in the low density regime to
use a hard sphere diameter in it. To rectify this defect, we replace the hard
sphere diameter $\sigma$ with the corresponding Chapman--Enskog collision
bracket integral for self-diffusion. Thus we insert the reduced
Chapman--Enskog collision bracket integral \cite{chapman} $\Omega^{(1,1)}$
into Eq. (\ref{10})and obtain the formula for the self-diffusion
coefficient\cite{qin} in the form%
\begin{equation}
D_{CE}=\frac{3}{8\rho\sigma^{2}\Omega^{(1,1)}}\sqrt{\frac{k_{B}T}{{\pi}m}}%
\exp(-{v^{c}}/{v_{f}}), \label{11}%
\end{equation}
where $\frac{\pi}{6}r^{\ddagger3}\equiv v^{c}$ as in Eq. (\ref{7}). This form
corrects the low density behavior.

\subsection{Model $D_{CERC}$}

This model $D_{CE}$ can be modified slightly if ${v^{c}}$ is replaced by
${v^{0}}$ introduced in Eq. (\ref{9}). We obtain\cite{jagtar}%
\begin{equation}
D_{CERC}=\frac{3}{8\rho\sigma^{2}\Omega^{(1,1)}}\sqrt{\frac{k_{B}T}{{\pi}m}%
}\exp(-{v^{0}}/{v_{f}}), \label{12}%
\end{equation}
where ${v^{0}}$ is defined by Eq. (\ref{8}) and $v_{f}$ by Eq. (\ref{5}).
Therefore the relation of $D_{CERC}$ to $D_{HSRC}$ is summarizable in the form%
\begin{equation}
D_{CERC}=\frac{D_{HSRC}}{\Omega^{(1,1)}}. \label{13}%
\end{equation}

The various models presented above are used to compute self-diffusion
coefficients and compared with the results by the molecular dynamics
simulation results\cite{md} $D_{MD}$. The results of comparison are presented
in the following Tables 1--6, where%
\begin{align*}
\Delta_{CERC}  &  =\frac{D_{CERC}-D_{MD}}{D_{MD}}\equiv\frac{D_{CC}-D_{MD}%
}{D_{MD}}=\Delta_{CC},\\
\Delta_{HSRC}  &  =\frac{D_{HSRC}-D_{MD}}{D_{MD}}\equiv\frac{D_{HC}-D_{MD}%
}{D_{MD}}=\Delta_{HC},\text{ etc.}%
\end{align*}
$D_{MD}$ denotes the molecular dynamics simulation value for the
self-diffusion coefficient. Self-diffusion coefficients $D_{CERC}$ and
$D_{HSRC}$ are abbreviated as $D_{HC}\equiv D_{HSRC}$ and $D_{CC}\equiv
D_{CERC}$ for the sake of formatting the Tables suitably.

Among the formulas presented, $D_{CERC}$ works best in the density regime of
$\rho^{\ast}\approx0.7$ with a maximum error of $4\%$, compared with the
MD\ simulation results. Therefore $D_{CERC}$ may be used to study related
properties of liquids or gases obeying the LJ interaction law. \ In the
density regime higher than $\rho^{\ast}\approx0.7$ none of the models performs
as well; for example, the errors of $D_{CERC}$ tend to be 20 to 40\% and the
errors of other models are comparable. In this regard, it should be noted that
the MD simulations results in the high density regime are rather small and
tend to be less reliable. At least, the models studied yield self-diffusion
coefficients of a correct order of magnitude and thus may be still useful
qualitative studies. \newpage

%

\begin{table}[tbp] \centering
\caption{\bf Self Diffusion Coefficient of the LJ fluid, {$\bf T^* = 6.0$}, {$\Omega^{(1,1) }= 0.8124$}}%
\begin{tabular}
[c]%
{p{0.8cm}p{0.8cm}p{0.8cm}p{0.8cm}p{0.8cm}p{0.8cm}p{0.8cm}p{0.8cm}p{0.8cm}p{0.8cm}}%
\hline
$\rho$ & $D_{MD}$ & $D_{CC}$ & $\Delta_{CC}$ & $D_{CE}$ & $\Delta_{CE}$ &
$D_{HC}$ & $\Delta_{HC}$ & $D_{HS}$ & $\Delta_{HS}$\\\hline
$0.025$ & $25.347$ & $25.349$ & $0.000$ & $25.214$ & $-0.005$ & $20.593$ &
$-0.188$ & $20.484$ & $-0.192$\\
$0.050$ & $12.487$ & $12.606$ & $0.010$ & $12.437$ & $-0.004$ & $10.241$ &
$-0.180$ & $10.103$ & $-0.191$\\
$0.075$ & $8.181$ & $8.324$ & $0.018$ & $8.097$ & $-0.010$ & $6.763$ &
$-0.173$ & $6.578$ & $-0.196$\\
$0.100$ & $6.016$ & $6.235$ & $0.036$ & $6.025$ & $0.001$ & $5.066$ & $-0.158$
& $4.895$ & $-0.186$\\
$0.150$ & $3.900$ & $4.108$ & $0.053$ & $3.873$ & $-0.007$ & $3.337$ &
$-0.144$ & $3.146$ & $-0.193$\\
$0.200$ & $2.850$ & $3.020$ & $0.060$ & $2.648$ & $-0.071$ & $2.454$ &
$-0.139$ & $2.151$ & $-0.245$\\
$0.250$ & $2.218$ & $2.370$ & $0.068$ & $2.005$ & $-0.096$ & $1.925$ &
$-0.132$ & $1.628$ & $-0.266$\\
$0.300$ & $1.793$ & $1.943$ & $0.084$ & $1.546$ & $-0.138$ & $1.579$ &
$-0.120$ & $1.256$ & $-0.300$\\
$0.350$ & $1.492$ & $1.629$ & $0.092$ & $1.194$ & $-0.200$ & $1.323$ &
$-0.113$ & $0.970$ & $-0.350$\\
$0.400$ & $1.267$ & $1.405$ & $0.109$ & $0.875$ & $-0.309$ & $1.142$ &
$-0.099$ & $0.711$ & $-0.439$\\
$0.450$ & $1.091$ & $1.211$ & $0.110$ & $0.724$ & $-0.336$ & $0.984$ &
$-0.098$ & $0.588$ & $-0.461$\\
$0.500$ & $0.953$ & $1.049$ & $0.101$ & $0.494$ & $-0.482$ & $0.853$ &
$-0.105$ & $0.401$ & $-0.579$\\
$0.550$ & $0.832$ & $0.936$ & $0.125$ & $0.405$ & $-0.513$ & $0.760$ &
$-0.086$ & $0.329$ & $-0.605$\\
$0.600$ & $0.729$ & $0.792$ & $0.086$ & $0.217$ & $-0.703$ & $0.643$ &
$-0.117$ & $0.176$ & $-0.758$\\
$0.650$ & $0.639$ & $0.715$ & $0.119$ & $0.154$ & $-0.758$ & $0.581$ &
$-0.091$ & $0.125$ & $-0.804$\\
$0.700$ & $0.564$ & $0.607$ & $0.077$ & $0.086$ & $-0.848$ & $0.493$ &
$-0.125$ & $0.070$ & $-0.876$\\
$0.750$ & $0.495$ & $0.558$ & $0.128$ & $0.071$ & $-0.856$ & $0.454$ &
$-0.083$ & $0.058$ & $-0.883$\\
$0.800$ & $0.433$ & $0.472$ & $0.090$ & $0.032$ & $-0.926$ & $0.383$ &
$-0.115$ & $0.026$ & $-0.940$\\
$0.850$ & $0.379$ & $0.446$ & $0.176$ & $0.022$ & $-0.941$ & $0.362$ &
$-0.045$ & $0.018$ & $-0.952$\\
$0.900$ & $0.329$ & $0.373$ & $0.133$ & $0.012$ & $-0.964$ & $0.303$ &
$-0.079$ & $0.010$ & $-0.970$\\
$0.950$ & $0.284$ & $0.296$ & $0.043$ & $0.002$ & $-0.994$ & $0.241$ &
$-0.152$ & $0.001$ & $-0.995$\\
$1.000$ & $0.244$ & $0.237$ & $-0.028$ & $0.000$ & $-0.998$ & $0.193$ &
$-0.210$ & $0.000$ & $-0.999$\\\hline
\end{tabular}
\label{table1}
\end{table}%
\newpage%
\begin{table}[tbp] \centering
\label{table2}
\caption{\bf Self Diffusion Coefficient of the LJ fluid, {$\bf T^* = 4.0$}, {$\Omega^{(1,1)} = 0.8836$}}%
\begin{tabular}
[c]%
{p{0.8cm}p{0.8cm}p{0.8cm}p{0.8cm}p{0.8cm}p{0.8cm}p{0.8cm}p{0.8cm}p{0.8cm}p{0.8cm}}%
\hline
$\rho$ & $D_{MD}$ & $D_{CC}$ & $r_{CC}$ & $D_{CE}$ & $r_{CE}$ & $D_{HC}$ &
$r_{HC}$ & $D_{HS}$ & $r_{HS}$\\\hline
$0.025$ & $18.911$ & $19.048$ & $0.007$ & $18.939$ & $0.001$ & $16.831$ &
$-0.110$ & $16.734$ & $-0.115$\\
$0.050$ & $9.315$ & $9.457$ & $0.015$ & $9.349$ & $0.004$ & $8.357$ & $-0.103$
& $8.261$ & $-0.113$\\
$0.075$ & $6.148$ & $6.263$ & $0.019$ & $6.128$ & $-0.003$ & $5.534$ &
$-0.100$ & $5.415$ & $-0.119$\\
$0.100$ & $4.557$ & $4.658$ & $0.022$ & $4.493$ & $-0.014$ & $4.116$ &
$-0.097$ & $3.970$ & $-0.129$\\
$0.150$ & $2.954$ & $3.045$ & $0.031$ & $2.850$ & $-0.035$ & $2.690$ &
$-0.089$ & $2.518$ & $-0.148$\\
$0.200$ & $2.160$ & $2.251$ & $0.042$ & $2.042$ & $-0.054$ & $1.989$ &
$-0.079$ & $1.805$ & $-0.165$\\
$0.250$ & $1.685$ & $1.752$ & $0.040$ & $1.498$ & $-0.111$ & $1.548$ &
$-0.081$ & $1.324$ & $-0.214$\\
$0.300$ & $1.363$ & $1.434$ & $0.052$ & $1.191$ & $-0.126$ & $1.267$ &
$-0.070$ & $1.052$ & $-0.228$\\
$0.350$ & $1.133$ & $1.195$ & $0.055$ & $0.877$ & $-0.226$ & $1.056$ &
$-0.068$ & $0.775$ & $-0.316$\\
$0.400$ & $0.960$ & $1.012$ & $0.054$ & $0.694$ & $-0.277$ & $0.894$ &
$-0.069$ & $0.614$ & $-0.361$\\
$0.450$ & $0.824$ & $0.860$ & $0.044$ & $0.515$ & $-0.375$ & $0.760$ &
$-0.077$ & $0.455$ & $-0.448$\\
$0.500$ & $0.710$ & $0.739$ & $0.040$ & $0.371$ & $-0.478$ & $0.653$ &
$-0.081$ & $0.327$ & $-0.539$\\
$0.550$ & $0.615$ & $0.646$ & $0.051$ & $0.286$ & $-0.535$ & $0.571$ &
$-0.072$ & $0.253$ & $-0.589$\\
$0.600$ & $0.536$ & $0.575$ & $0.073$ & $0.226$ & $-0.579$ & $0.508$ &
$-0.052$ & $0.199$ & $-0.628$\\
$0.650$ & $0.464$ & $0.500$ & $0.077$ & $0.152$ & $-0.671$ & $0.442$ &
$-0.048$ & $0.135$ & $-0.710$\\
$0.700$ & $0.403$ & $0.441$ & $0.095$ & $0.114$ & $-0.717$ & $0.390$ &
$-0.032$ & $0.101$ & $-0.750$\\
$0.800$ & $0.300$ & $0.314$ & $0.045$ & $0.036$ & $-0.878$ & $0.277$ &
$-0.076$ & $0.032$ & $-0.893$\\
$0.850$ & $0.256$ & $0.248$ & $-0.032$ & $0.014$ & $-0.947$ & $0.219$ &
$-0.145$ & $0.012$ & $-0.953$\\
$0.900$ & $0.217$ & $0.220$ & $0.013$ & $0.010$ & $-0.952$ & $0.194$ &
$-0.105$ & $0.009$ & $-0.958$\\
$0.950$ & $0.183$ & $0.176$ & $-0.036$ & $0.004$ & $-0.980$ & $0.156$ &
$-0.149$ & $0.003$ & $-0.982$\\
$1.000$ & $0.153$ & $0.144$ & $-0.058$ & $0.002$ & $-0.987$ & $0.127$ &
$-0.167$ & $0.002$ & $-0.989$\\\hline
\end{tabular}%
\end{table}%
\newpage%

\begin{table}[tbp] \centering
\caption{\bf Self Diffusion Coefficient of the LJ fluid, {$\bf T^* = 1.3$}, {$\Omega^{(1,1)} = 1.273$}}%
\begin{tabular}
[c]%
{p{0.8cm}p{0.8cm}p{0.8cm}p{0.8cm}p{0.8cm}p{0.8cm}p{0.8cm}p{0.8cm}p{0.8cm}p{0.8cm}}%
\hline
$\rho$ & $D_{MD}$ & $D_{CC}$ & $r_{CC}$ & $D_{CE}$ & $r_{CE}$ & $D_{HC}$ &
$r_{HC}$ & $D_{HS}$ & $r_{HS}$\\\hline
$0.025$ & $7.462$ & $7.515$ & $0.007$ & $7.473$ & $0.001$ & $9.566$ & $0.282$
& $9.512$ & $0.275$\\
$0.050$ & $3.684$ & $3.726$ & $0.011$ & $3.678$ & $-0.002$ & $4.744$ & $0.288$
& $4.683$ & $0.271$\\
$0.075$ & $2.449$ & $2.449$ & $0.000$ & $2.409$ & $-0.016$ & $3.118$ & $0.273$
& $3.067$ & $0.252$\\
$0.100$ & $1.843$ & $1.818$ & $-0.014$ & $1.774$ & $-0.037$ & $2.314$ &
$0.256$ & $2.258$ & $0.225$\\
$0.150$ & $1.229$ & $1.182$ & $-0.039$ & $1.139$ & $-0.074$ & $1.504$ &
$0.224$ & $1.450$ & $0.179$\\
$0.200$ & $0.924$ & $0.855$ & $-0.074$ & $0.802$ & $-0.132$ & $1.089$ &
$0.178$ & $1.021$ & $0.105$\\
$0.450$ & $0.353$ & $0.307$ & $-0.130$ & $0.246$ & $-0.302$ & $0.391$ &
$0.107$ & $0.314$ & $-0.111$\\
$0.500$ & $0.299$ & $0.260$ & $-0.131$ & $0.203$ & $-0.322$ & $0.331$ &
$0.106$ & $0.258$ & $-0.137$\\
$0.550$ & $0.254$ & $0.216$ & $-0.151$ & $0.159$ & $-0.373$ & $0.275$ &
$0.081$ & $0.203$ & $-0.202$\\
$0.600$ & $0.212$ & $0.191$ & $-0.100$ & $0.136$ & $-0.359$ & $0.243$ &
$0.146$ & $0.173$ & $-0.184$\\
$0.650$ & $0.178$ & $0.159$ & $-0.106$ & $0.105$ & $-0.407$ & $0.203$ &
$0.139$ & $0.134$ & $-0.246$\\
$0.700$ & $0.145$ & $0.128$ & $-0.119$ & $0.076$ & $-0.478$ & $0.163$ &
$0.121$ & $0.096$ & $-0.335$\\
$0.750$ & $0.117$ & $0.100$ & $-0.147$ & $0.053$ & $-0.549$ & $0.127$ &
$0.086$ & $0.067$ & $-0.425$\\
$0.800$ & $0.094$ & $0.075$ & $-0.204$ & $0.035$ & $-0.631$ & $0.095$ &
$0.013$ & $0.044$ & $-0.530$\\
$0.850$ & $0.073$ & $0.059$ & $-0.197$ & $0.024$ & $-0.677$ & $0.075$ &
$0.022$ & $0.030$ & $-0.589$\\
$0.900$ & $0.055$ & $0.035$ & $-0.362$ & $0.011$ & $-0.795$ & $0.045$ &
$-0.187$ & $0.014$ & $-0.739$\\
$0.950$ & $0.040$ & $0.033$ & $-0.184$ & $0.009$ & $-0.766$ & $0.042$ &
$0.039$ & $0.012$ & $-0.702$\\\hline
\end{tabular}
\label{table3}
\end{table}%
\newpage%
\begin{table}[tbp] \centering
\caption{\bf Self Diffusion Coefficient of the LJ fluid, {$\bf T^* = 1.0$}, {$\Omega^{(1,1)} = 1.439$}}%
\begin{tabular}
[c]%
{p{0.8cm}p{0.8cm}p{0.8cm}p{0.8cm}p{0.8cm}p{0.8cm}p{0.8cm}p{0.8cm}p{0.8cm}p{0.8cm}}%
\hline
$\rho$ & $D_{MD}$ & $D_{CC}$ & $r_{CC}$ & $D_{CE}$ & $r_{CE}$ & $D_{HC}$ &
$r_{HC}$ & $D_{HS}$ & $r_{HS}$\\\hline
$0.005$ & $29.427$ & $29.354$ & $-0.002$ & $29.328$ & $-0.003$ & $42.241$ &
$0.435$ & $42.202$ & $0.434$\\
$0.010$ & $14.547$ & $14.649$ & $0.007$ & $14.627$ & $0.005$ & $21.080$ &
$0.449$ & $21.048$ & $0.447$\\
$0.015$ & $9.595$ & $9.744$ & $0.016$ & $9.718$ & $0.013$ & $14.022$ & $0.461$
& $13.985$ & $0.458$\\
$0.020$ & $7.203$ & $7.290$ & $0.012$ & $7.267$ & $0.009$ & $10.490$ & $0.456$
& $10.457$ & $0.452$\\
$0.025$ & $5.710$ & $5.825$ & $0.020$ & $5.797$ & $0.015$ & $8.382$ & $0.468$
& $8.342$ & $0.461$\\
$0.030$ & $4.770$ & $4.842$ & $0.015$ & $4.818$ & $0.010$ & $6.967$ & $0.461$
& $6.933$ & $0.453$\\
$0.035$ & $4.040$ & $4.140$ & $0.025$ & $4.113$ & $0.018$ & $5.957$ & $0.475$
& $5.919$ & $0.465$\\
$0.040$ & $3.529$ & $3.613$ & $0.024$ & $3.584$ & $0.016$ & $5.200$ & $0.473$
& $5.158$ & $0.461$\\
$0.045$ & $3.131$ & $3.210$ & $0.025$ & $3.183$ & $0.016$ & $4.619$ & $0.475$
& $4.580$ & $0.463$\\
$0.050$ & $2.813$ & $2.876$ & $0.022$ & $2.849$ & $0.013$ & $4.138$ & $0.471$
& $4.100$ & $0.457$\\
$0.055$ & $2.544$ & $2.608$ & $0.025$ & $2.579$ & $0.014$ & $3.752$ & $0.475$
& $3.711$ & $0.459$\\
$0.060$ & $2.348$ & $2.392$ & $0.019$ & $2.363$ & $0.006$ & $3.443$ & $0.466$
& $3.400$ & $0.448$\\
$0.065$ & $2.169$ & $2.198$ & $0.014$ & $2.168$ & $-0.001$ & $3.163$ & $0.458$
& $3.119$ & $0.438$\\
$0.070$ & $2.014$ & $2.033$ & $0.010$ & $2.002$ & $-0.006$ & $2.926$ & $0.453$
& $2.881$ & $0.431$\\
$0.650$ & $0.139$ & $0.119$ & $-0.147$ & $0.092$ & $-0.339$ & $0.171$ &
$0.228$ & $0.132$ & $-0.050$\\
$0.700$ & $0.111$ & $0.095$ & $-0.145$ & $0.068$ & $-0.384$ & $0.137$ &
$0.230$ & $0.098$ & $-0.113$\\
$0.750$ & $0.089$ & $0.079$ & $-0.109$ & $0.057$ & $-0.365$ & $0.114$ &
$0.281$ & $0.081$ & $-0.086$\\
$0.800$ & $0.068$ & $0.052$ & $-0.235$ & $0.034$ & $-0.505$ & $0.075$ &
$0.101$ & $0.048$ & $-0.287$\\
$0.850$ & $0.051$ & $0.035$ & $-0.313$ & $0.021$ & $-0.584$ & $0.050$ &
$-0.011$ & $0.031$ & $-0.401$\\
$0.900$ & $0.037$ & $0.027$ & $-0.258$ & $0.017$ & $-0.553$ & $0.040$ &
$0.068$ & $0.024$ & $-0.357$\\\hline
\end{tabular}
\label{table4}
\end{table}%
\newpage%
\begin{table}[tbp] \centering
\caption{\bf Self Diffusion Coefficient of the LJ fluid, {$\bf T^* = 0.8$}, {$\Omega^{(1,1) }= 1.612$}}%
\begin{tabular}
[c]%
{p{0.8cm}p{0.8cm}p{0.8cm}p{0.8cm}p{0.8cm}p{0.8cm}p{0.8cm}p{0.8cm}p{0.8cm}p{0.8cm}}%
\hline
$\rho$ & $D_{MD}$ & $D_{CC}$ & $r_{CC}$ & $D_{CE}$ & $r_{CE}$ & $D_{HC}$ &
$r_{HC}$ & $D_{HS}$ & $r_{HS}$\\\hline
$0.005$ & $23.331$ & $23.436$ & $0.004$ & $23.415$ & $0.004$ & $37.778$ &
$0.619$ & $37.744$ & $0.618$\\
$0.010$ & $11.455$ & $11.692$ & $0.021$ & $11.676$ & $0.019$ & $18.847$ &
$0.645$ & $18.822$ & $0.643$\\
$0.015$ & $7.589$ & $7.780$ & $0.025$ & $7.759$ & $0.022$ & $12.542$ & $0.653$
& $12.508$ & $0.648$\\
$0.020$ & $5.646$ & $5.817$ & $0.030$ & $5.797$ & $0.027$ & $9.378$ & $0.661$
& $9.345$ & $0.655$\\
$0.025$ & $4.460$ & $4.644$ & $0.041$ & $4.622$ & $0.036$ & $7.486$ & $0.679$
& $7.450$ & $0.671$\\
$0.030$ & $3.709$ & $3.862$ & $0.041$ & $3.841$ & $0.036$ & $6.225$ & $0.678$
& $6.192$ & $0.670$\\
$0.700$ & $0.088$ & $0.070$ & $-0.206$ & $0.059$ & $-0.331$ & $0.113$ &
$0.280$ & $0.095$ & $0.079$\\
$0.750$ & $0.069$ & $0.047$ & $-0.316$ & $0.040$ & $-0.423$ & $0.076$ &
$0.102$ & $0.064$ & $-0.070$\\
$0.800$ & $0.051$ & $0.039$ & $-0.234$ & $0.035$ & $-0.313$ & $0.063$ &
$0.235$ & $0.056$ & $0.107$\\
$0.850$ & $0.037$ & $0.024$ & $-0.345$ & $0.020$ & $-0.466$ & $0.039$ &
$0.056$ & $0.032$ & $-0.138$\\\hline
\end{tabular}
\label{table5}
\end{table}%
\newpage%
\begin{table}[tbp] \centering
\caption{\bf Self Diffusion Coefficient of the LJ fluid, {$\bf T^* = 0.7$}, {$\Omega^{(1,1)} = 1.729$}}%
\begin{tabular}
[c]%
{p{0.8cm}p{0.8cm}p{0.8cm}p{0.8cm}p{0.8cm}p{0.8cm}p{0.8cm}p{0.8cm}p{0.8cm}p{0.8cm}}%
\hline
$\rho$ & $D_{MD}$ & $D_{CC}$ & $r_{CC}$ & $D_{CE}$ & $r_{CE}$ & $D_{HC}$ &
$r_{HC}$ & $D_{HS}$ & $r_{HS}$\\\hline
$0.005$ & $20.223$ & $20.433$ & $0.010$ & $20.418$ & $0.010$ & $35.329$ &
$0.747$ & $35.303$ & $0.746$\\
$0.010$ & $9.979$ & $10.194$ & $0.022$ & $10.179$ & $0.020$ & $17.625$ &
$0.766$ & $17.599$ & $0.764$\\
$0.015$ & $6.562$ & $6.785$ & $0.034$ & $6.768$ & $0.031$ & $11.732$ & $0.788$
& $11.702$ & $0.783$\\
$0.750$ & $0.058$ & $0.042$ & $-0.275$ & $0.040$ & $-0.311$ & $0.073$ &
$0.253$ & $0.069$ & $0.191$\\
$0.800$ & $0.043$ & $0.029$ & $-0.322$ & $0.029$ & $-0.321$ & $0.050$ &
$0.173$ & $0.050$ & $0.174$\\
$0.850$ & $0.030$ & $0.015$ & $-0.494$ & $0.016$ & $-0.474$ & $0.026$ &
$-0.125$ & $0.027$ & $-0.090$\\\hline
\end{tabular}
\label{table6}
\end{table}%


\newpage

\begin{thebibliography}{9}                                                                                                %


\bibitem {mfv}K. Rah and B. C. Eu, J. Chem. Phys. \textbf{115}, 2634 (2001);
Phys. Rev. Lett. \textbf{83}, 4566 (1999); Phys. Rev. E \textbf{60}, 4105
(1999); \textbf{116}, 7967 (2002).

\bibitem {GvdW}B. C. Eu and K. Rah, {Phys. Rev. E.} \textbf{63}, 031203
(2001); B. C. Eu, J. Chem. Phys. \textbf{114}, 10899 (2001); K. Rah and B. C.
Eu, J. Phys. Chem. B \textbf{107}, 4382 (2003).

\bibitem {laghaei}R. Laghaei, A. E. Eskandari,and B. C. Eu, J. Phys. Chem. B
\textbf{109}, 5873--5883 (2005)

\bibitem {excluded}R. Laghaei, A. E. Eskandari,and B. C. Eu, {J. Chem. Phys.}
\textbf{124}, 154502 (2006).

\bibitem {chapman}S. Chapman and T. G. Cowling, \textit{The Mathematical
Theory of Nonuniform Gases} (Cambridge U.P., London, 1970), third ed. For
tables tabulated for $\Omega^{(1,1)}$ for the LJ fluids, see J. O.
Hirschfelder, C. F. Curtis, and R. B. Bird, \textit{Molecular Theory of Gases
and Liquids (}Wiley, New York, 1954), Appendix, Table I-M\textbf{.}

\bibitem {qin}Y. Qin and B. C. Eu., {J. Phys. Chem. B.} \textbf{113}, 4751 (2009).

\bibitem {jagtar}J. S. Singh and B. C. Eu (to be submitted).

\bibitem {md}K. Meier, A. Laesecke, and S. Kabalac, J. Chem. Phys.
\textbf{121}, 9526 (2004).
\end{thebibliography}
\end{document}